\newcommand{\lsim}{\alt}
\newcommand{\gsim}{\agt}

\newcommand{\hi}{\hat{\imath}}
\newcommand{\hj}{\hat{\jmath}}
\newcommand{\labx}[1]{\label{#1}}

\documentclass[prb%
,a4paper%
,aps%
,twocolumn%
,amssymb%
,preprintnumbers%
,nofootinbib%
,superscriptaddress%
,amsmath%
]{revtex4}
\usepackage{hyperref}
\usepackage{epsfig}
\begin{document}

\title{Spontaneous vortex formation on a superconductor film}

\author{M.~Donaire}
\affiliation{DAMTP, CMS, University of Cambridge,
Wilberforce Road,\\
Cambridge CB3 0WA, United Kingdom
}

\author{T.W.B.~Kibble}
\affiliation{Blackett Laboratory, Imperial
College London,\\London SW7 2AZ, United Kingdom}

\author{A.~Rajantie}
\affiliation{DAMTP, CMS, University of Cambridge,
Wilberforce Road,\\
Cambridge CB3 0WA, United Kingdom
}

\date{September 7, 2004}

\begin{abstract}
We carry out numerical simulations to
investigate spontaneous vortex formation
during a temperature quench of a superconductor film
from the normal to the superconducting phase in the absence of
an external magnetic field. Our results agree roughly with 
quantitative predictions
of the flux trapping scenario: In fast quenches the
agreement is almost perfect, but there appears to be some discrepancy in
slower ones. In particular, our simulations
demonstrate the crucial role the electromagnetic field plays in this 
phenomenon, making it very different from vortex formation in superfluids.
Besides superconductor experiments, our findings also shed more light
on the possible formation of cosmic defects in the early universe.
\end{abstract}

\preprint{DAMTP-2004-88, Imperial/TP/040905}
\preprint{cond-mat/0409172}
\maketitle

\section{Introduction}

When a superconductor in rapidly cooled from the normal to the superconducting
phase in the absence of an external magnetic field, vortices (and antivortices)
form. This phenomenon has been observed in several 
experiments\cite{%
carmi:7595,%
carmi:4966,%
Kavoussanaki:2000tj,%
monaco:080603,%
monaco:104506,%
kirtley:257001,%
maniv:197001,%
Kir04%
}
and has analogues in other condensed matter systems such as 
superfluids\cite{Bau+96,Ruutu:1996qz,Dod+98} and liquid 
crystals.\cite{Chu+91,Bow+94,Dig98,Ray04}
In the latter cases, the order parameter is electrically neutral and 
defect formation can be understood in terms of the
Kibble-Zurek mechanism.\cite{Kibble:1976sj,Kibble:1980mv,Zurek:1996sj} 
In contrast, the Cooper pairs in superconductors are electrically 
charged, which means that the electromagnetic field plays an
important role.\cite{Hindmarsh:2000kd,Rajantie:2001ps}

Understanding the mechanisms in which vortices form would also have 
implications well beyond condensed matter physics. In particular, 
similar mechanisms may have produced topological defects such as cosmic 
strings, magnetic monopoles and domain walls in the early 
universe,\cite{Kibble:1976sj,VilShel} either
in cosmological phase transitions\cite{Kibble:1980mv}
or in brane collisions.\cite{Dvali:2003zj}
This is particularly interesting, because
two possible gravitational lensing events
by cosmic strings have recently been 
observed.\cite{Sazhin:2003cp,Schild:2004uv,Sazhin:2004fv}

In cosmology, a neutral order parameter corresponds to a global symmetry and
a charged order parameter to a local gauge symmetry. In the latter
case, which is more common,
the vortices are Nielsen-Olesen cosmic strings.\cite{Nielsen:1973cs}
 The role of the
Cooper pairs is played by the Higgs field, which obtains a non-zero
vacuum expectation value at the transition.

In this paper, we investigate vortex formation in the case of a 
two-dimensional film in three-dimensional space. This is the relevant 
setup for
actual superconductor experiments, but to our knowledge no simulations
have been carried out in the past. While in the case of a neutral order
parameter, a two-dimensional film can be studied using a fully
two-dimensional simulation, the three-dimensional nature of the
electromagnetic field means that the same is not true for
superconductors.\cite{Kib03}

Because our setup captures some of the pecularities of fully three-dimensional
systems, it is also more relevant for cosmological applications than previous
two-dimensional studies.\cite{Stephens:2001fv}
A different borderline case between two and three spatial dimensions
has been studied in the past in 
Refs.~\onlinecite{Hindmarsh:2000kd} and \onlinecite{Hindmarsh:2001vp}.

\section{Setup}
The system we shall study consists of a two-dimensional superconductor
film in a three-dimensional space. The superconductor is described by the
Ginzburg-Landau theory with an electrically charged order parameter that 
is confined to the film. In contrast, the electromagnetic field lives
in the full three-dimensional space.
We shall denote three-vectors by bold-faced letters (such as 
$\mathbf{x}$) and two-vectors on the film by letters with arrows on top
($\vec{x}$).

We consider a hypothetical experiment in which the whole system is first 
heated up to a temperature above $T_c$, so that the film is in the normal 
phase. Then the system is gradually cooled through the phase transition
to the superconducting phase.

As usual, the two-dimensional Ginzburg-Landau (free) energy
is written as
\begin{equation}
E_\psi = \int d^2x\left[
\alpha |\psi|^2 + \frac{\beta}{2} |\psi|^4 + \frac{1}{2m} 
\left| \left(-i\hbar\vec\nabla - e\vec{A} \right) \psi \right|^2
\right]\!.
\labx{equ:Epsi}
\end{equation}
The field $\psi(x,y)$ is a complex scalar field, and the two-dimensional
vector potential $\vec{A}(x,y)$ is obtained from the full three-dimensional
$\mathbf{A}(x,y,z)$ 
by restricting to the plane $z=0$, and $x$ and $y$ directions,
\begin{equation}
\vec{A}_x(x,y)=\mathbf{A}_x(x,y,0),\quad
\vec{A}_y(x,y)=\mathbf{A}_y(x,y,0).
\end{equation}
The energy of a three-dimensional electromagnetic field 
configuration is given by
\begin{equation}
E_{\mathbf{A}}=\int d^3x\left[\frac{\epsilon_0\mathbf{E}^2}{2}+
\frac{\mathbf{B}^2}{2\mu_0}\right],
\labx{equ:EA}
\end{equation}
where $\mathbf{B}=\mathbf{\nabla}\times\mathbf{A}$ is the magnetic 
field (or induction) and $\mathbf{E}$ is the electric field. If we choose the
temporal gauge, we can write $\mathbf{E}=-\partial\mathbf{A}/\partial t$.
In what follows, we shall simplify the notation by choosing a unit system in
which $m=1/2$, $c=\hbar=\mu_0=\epsilon_0=k_B=1$.

Because we are interested in the time evolution of the system, we
need to include the kinetic energy term corresponding to the time derivative
$\pi=\dot\psi$, for which we choose the standard ``relativistic'' form.
The total energy is then given by
\begin{eqnarray}
E_{\rm tot} &=&
\int d^2x\left[
|\pi|^2+\alpha |\psi|^2 + \frac{\beta}{2} |\psi|^4 +
| \vec{D} \psi |^2\right]\nonumber\\
&&+\frac{1}{2}\int d^3x\left(\mathbf{E}^2+
\mathbf{B}^2
\right),
\labx{equ:totalenergy}
\end{eqnarray}
where we have used the covariant derivative $\vec{D}\psi=(\vec\nabla
+ie\vec{A})\psi$.
The full partition function is
\begin{equation}
Z=\int {\cal D}\psi{\cal D}\pi{\cal D}\mathbf{A}{\cal D}\mathbf{E}\,
e^{-E_{\rm tot}/T}.
\labx{equ:pf}
\end{equation}
This partition function describes the thermal equilibrium state of the system.
We shall use the thermal ensemble as the ensemble of 
initial conditions for the time evolution, choosing a high enough temperature
so that the film in initially in the normal phase.

The equations of motion are obtained by using $E_{\rm tot}$ as the Hamiltonian 
and identifying $\pi^*$ as the canonical momentum conjugate to
$\psi$. This gives the equations
\begin{eqnarray}
\ddot\psi-\vec{D}^2\psi+\alpha\psi+\beta|\psi|^2\psi&=&0,\nonumber\\
\ddot{\mathbf A}+\mathbf{\nabla}\times\mathbf{\nabla}\times\mathbf{A}
&=&\mathbf{j},
\labx{equ:eom0}
\end{eqnarray}
where $\mathbf{j}$
is the electric current. 
The latter 
equation is, of course, equivalent to Maxwell's equations.
The current is confined on the plane $z=0$, i.e.,
$\mathbf{j}=\delta(z)(\vec{j}_x,\vec{j}_y,0)$, where the two-dimensional
current is $\vec{j}=2e{\rm Im}\psi^*\vec{D}\psi$.

Because Eq.~(\ref{equ:pf}) describes a thermal equilibrium state, 
the system stays in equilibrium if it is evolved according to 
Eqs.~(\ref{equ:eom0}). 
In order to cool the system, we modify the equations by adding a 
damping term to the equation for $\psi$,
\begin{equation}
\ddot\psi-\vec{D}^2\psi+\alpha\psi+\beta|\psi|^2\psi=-\sigma\dot\psi.
\labx{equ:eom}
\end{equation}
In order not to violate Gauss's law $\mathbf{\nabla}\cdot\mathbf{E}=\rho$,
we need to modify the equation for the vector potential $\mathbf{A}$ as well,
by adding an Ohmic contribution to the electric current,
\begin{equation}
\mathbf{j}=\delta(z)(\vec{j}_x,\vec{j}_y,0)+\sigma\mathbf{E},
\end{equation}
where the damping rate $\sigma$ plays the role of the conductivity.

When the system is evolved according to these equations, the total energy 
gradually decreases. Strictly speaking, the system will not be in thermal
equilibrium, but using a suitable effective definition of temperature, one
can say that it cools down. If $\alpha$ is negative, the system will
eventually reach the critical temperature at which the film becomes
superconducting. 

Note that Eq.~(\ref{equ:eom}), with an appropriate rescaling, 
becomes the usual time-dependent Ginzburg-Landau
equation in the limit $\sigma\rightarrow\infty$. 
However, $\sigma$ determines
the rate of cooling of the system, which we want to be able to vary. Therefore,
we have to keep the second-order terms both in Eq.~(\ref{equ:eom}) and in
the Maxwell equation.

A damping term like this is not the only possible way to induce a phase
transition. In Ref.~\onlinecite{Hindmarsh:2000kd}, this was done by varying 
the quadratic
term $\alpha$ at constant temperature. We do not believe the qualitative
phenomena we
are interested in are sensitive to the specific choice, but quantitative
details may well be different.

\section{Flux trapping}
As long as the system is in the normal phase, it will stay close to
thermal equilibrium, because the film does not affect the dynamics of
the vector potential significantly. To a good approcimation, 
the Fourier modes of the orthogonal
magnetic field $B=\partial_x \vec{A}_y-\partial_y \vec{A}_x$ with
wave number $\vec{k}$ greater than $\sigma$ will 
oscillate with a decreasing amplitude
\begin{equation}
B(\vec{k})\sim\exp(-\sigma t/2)\cos(kt).
\end{equation}
The equilibrium distribution given by Eq.~(\ref{equ:pf}) 
is approximately Gaussian
in the normal phase, and it can therefore be completely characterized by
the two-point function
\begin{equation}
\langle B(\vec{k}) B(\vec{k}')\rangle\equiv
G(|\vec{k}|)(2\pi)^2\delta(\vec{k}+\vec{k}'),
\end{equation}
where\cite{Kib03}
\begin{equation}
G(k)\approx \frac{Tk}{2}.
\end{equation}
Thus, the modes with $|\vec{k}|\ll\sigma$ retain 
the equilibrium distribution with an exponentially 
decreasing effective 
temperature $T_{\rm eff}(t)\approx T_{\rm ini}\exp(-\sigma t)$.

When the temperature reaches the critical value $T_c$, the 
order parameter $\psi$ becomes non-zero and the dynamics becomes non-linear.
If the system were to stay
in equilibrium, it would now be in the superconducting phase and therefore
repel magnetic fields. 

The equilibrium two-point function is\cite{Kib03}
\begin{equation}
G(k)\approx \frac{Tk}{2}\frac{1}{1+k\Lambda},
\end{equation}
where $\Lambda$ is a screening length that starts from zero at $T_c$
and grows exponentially as the temperature decreases.
At any finite temperature, thermal fluctuations with wavelength less than
$\Lambda$ are suppressed relative to the normal phase, but longer wavelengths 
are unaffected. 
This means that there is no sharp transition but the apparent
critical temperature depends on the length scale.

Even though it is impossible to solve the non-linear 
equations of motion analytically, 
we can generally say that the amplitude of
a Fourier mode of $B$ with wave number $k$ cannot
decay arbitrarily fast. Moreover, the longer wave lengths react slower.
For instance, if the dynamics of the long-wavelength fluctuations is
diffusive, the fastest possible decay rate is 
$\gamma_{\rm max}(k)=k^2/D$, where $D$ is the diffusion constant. 
It is therefore unavoidable that if the transition takes place in a 
finite time, the longest wavelengths are too slow to react, and
freeze out. This means that there is a critical wave number $k_c$
so that modes with $k<k_c$ still have approximately their initial
amplitude after the transition. 

The survival of the long-wavelength fluctuations means that at distances less
than $1/k_c$, there is effectively a uniform magnetic field. If this length
scale is longer than the Pearl length, i.e., the size of a vortex, 
this magnetic field must form an Abrikosov vortex 
lattice.\cite{Hindmarsh:2000kd} 
This mechanism of vortex formation is called flux trapping.

The number density of vortices per unit area
produced by flux trapping
is approximately\cite{Kib03}
\begin{equation}
n\approx\frac{e}{2\pi}\sqrt{\frac{T_ck_c^3}{2\pi}}.
\labx{equ:npred}
\end{equation}
Moreover, because the frozen-out magnetic field consists of modes with
wavelengths longer than $2\pi/k_c$, there are clusters of size $2\pi/k_c$
of equal-sign vortices. We can estimate the number of vortices
in each cluster by assuming that they are disks of radius $1/k_c$.
\begin{equation}
N_{\rm cl}\approx \sqrt{\frac{e^2T_c}{8\pi k_c}}
\labx{equ:Nclpred}
\end{equation}
vortices. 
For the estimate in Eq.~(\ref{equ:npred}) to be valid, $N_{\rm cl}$ has
to be greater than one. Otherwise, one will have to consider the coupled
dynamics of both the electromagnetic field and the order parameter.
It seems likely that vortex formation can
then be described by the Kibble-Zurek 
mechanism,\cite{Kibble:1976sj,Kibble:1980mv,Zurek:1996sj}
although the electromagnetic field may still play a role.

In fact, the friction term $\sigma$ causes 
modes with $|\vec{k}|\lsim\sigma$ to freeze out
even in the absence of any critical
dynamics. The solution of the linearized equations of motion with
thermal initial conditions gives
\begin{eqnarray}
G(k) &=& Te^{-\sigma t}\!
\int\! \frac{dk_z}{2\pi}
\frac{\vec{k}^2}{\mathbf k^2}
\left[
\frac{\sigma^2}{\tilde\sigma^2}\cosh\tilde\sigma t
+\frac{\sigma}{\tilde\sigma}\sinh\tilde\sigma t
-\frac{4\mathbf k^2}{\tilde\sigma^2}
\right]\nonumber\\
&\approx&
T\!
\int\! \frac{dk_z}{2\pi}
\frac{\vec{k}^2}{\mathbf k^2}
e^{-(2\mathbf k^2/\sigma)t}
=\frac{Tk}{2}\textrm{Erfc}\sqrt{2k^2t/\sigma}
,
\labx{equ:exact2p}
\end{eqnarray}
where $\mathbf k^2=\vec{k}^2+k_z^2$ and $\tilde\sigma^2=\sigma^2
-4\mathbf k^2$,
and $\textrm{Erfc}$ is the complementary error integral.
The second line is valid at late times when $|\vec{k}|\ll\sigma$.
This gives rise to an apparent critical wave number $k_c=\sqrt{\sigma/2t}$.

We are mainly interested in the physically more relevant
case in which the freeze-out is caused
by the critical dynamics, which restricts us to relatively low damping
rates $\sigma$. Nevertheless, simulations with higher $\sigma$
are also useful, because we can make more precise
theoretical predictions using the exact two-point function
(\ref{equ:exact2p}).

\section{Flux quantization}

Let us now assume that at the freeze-out, the two-point function
is given by some function $G(k)$, and calculate how many vortices should be
formed.
If the amplitude of the frozen-out fluctuations is high enough,
this can be done 
by considering a circular
region of radius $R$. 
We assume that the magnetic field is more or less uniform at this length scale
and that the magnetic flux $\Phi(R)$ through this region is much greater than 
one flux quantum $2\pi/e$. In that case, the number of vortices
$N(R)$ is given by the flux
\begin{equation}
N(R)=n\pi R^2=\frac{e}{2\pi}|\Phi(R)|
=\frac{e}{2\pi}\left|\int_0^R d^2x B(\vec{x})\right|,
\label{equ:NR}
\end{equation}
where $n$ is the number density of vortices per unit area.
We can therefore write
\begin{equation}
n^2=\lim_{R\rightarrow 0}\frac{e^2}{4\pi^4 R^4}\langle \Phi(R)^2\rangle.
\end{equation}
It is straightforward to write $\langle \Phi(R)^2\rangle$
in terms of $G(k)$,
\begin{eqnarray}
\langle \Phi(R)^2\rangle
&=&\int d^2x d^2y
\langle B(\vec{x})B(\vec{y})\rangle
\nonumber\\&=&
R^4\int d^2k\left(\frac{J_1(kR)}{kR}\right)^2G(k)
\nonumber\\&\xrightarrow{R\rightarrow 0}& 
\frac{R^4}{4}\int d^2kG(k)
,
\end{eqnarray}
where $J_1(kR)$ is a Bessel function.
Thus, we expect
\begin{equation}
n=\frac{e}{4\pi^2}\left(\int d^2kG(k)\right)^{1/2}.
\labx{equ:npredfull}
\end{equation}

In the friction-dominated case, we can combine the integrations in 
Eqs.~(\ref{equ:exact2p}) and (\ref{equ:npredfull}) into one 
integral,
\begin{eqnarray}
n^2
\!&=&\!\frac{e^2Te^{-\sigma t}}{12\pi^4}
\int d\mathbf k \mathbf k^2
\left[
\frac{\sigma^2}{\tilde\sigma^2}\cosh\tilde\sigma t
+\frac{\sigma}{\tilde\sigma}\sinh\tilde\sigma t
-\frac{4\mathbf k^2}{\tilde\sigma^2}
\right]\nonumber\\
&\approx&
\frac{e^2T}{12\pi^4}
\int d\mathbf k \mathbf k^2
e^{-2\mathbf k^2t/\sigma}
=\frac{e^2T}{48\pi^2}\left(
\frac{\sigma}{2\pi t}
\right)^{3/2}
,
\labx{equ:npredfast}
\end{eqnarray}
where the approximate equality is valid at late times.
In principle, $t$ should be chosen to be the time the vortices form.

For slower $\sigma$, the two-point function $G(k)$ cannot be
calculated analytically, but we assume that it can still be
approximated by a
Gaussian function,
\begin{equation}
G(k)=\frac{T_2k}{2}e^{-(k/k_2)^2},
\labx{equ:fitGaussian}
\end{equation}
where $T_2$ and $k_2$ are constants.
Then, Eq.~(\ref{equ:npredfull}) yields
\begin{equation}
n^2=\frac{e^2T_2k_2^3}{64\pi^{5/2}}.
\labx{equ:npredslow}
\end{equation}
The number of vortices per cluster is roughly given by $N_{\rm cl}\approx
n\pi/k_c^2$,
which gives
\begin{equation}
N_{\rm cl}\approx \sqrt{\frac{e^2T_2}{64\pi^{1/2}k_2}}
,\label{equ:Ncl}
\end{equation}
and in principle this number should be much greater than one for 
Eq.~(\ref{equ:NR}), and consequently also these
estimates, to be valid.

It is important to note that
Eqs.~(\ref{equ:exact2p}) and (\ref{equ:fitGaussian}) are idealized
descriptions of what the two-point function $G(k)$ should look like at the
time of the freeze-out, when the magnetic field is still smooth
over long distances.
When the vortices form, the magnetic flux is quantized and this introduces
microscopic structure to the two-point function. This will be important
when we measure it and attempt to extract the fit parameters.

Let us first assume that $N$ vortices and $N$ antivortices are formed,
and that they are point-like so that their magnetic field can be 
described by a sum of delta funtions,
\begin{equation}
B(\vec{x})=\frac{2\pi}{e} 
\sum_{i=1}^N \left[ \delta(x-x^+_i) - \delta(x-x^-_i) \right],
\end{equation}
where $\vec{x}^\pm_i$ are the positions of the vortices and antivortices.
The two-point function is
\begin{equation}
\langle B(\vec{x})B(\vec{y})\rangle 
= \frac{4\pi^2}{e^2} \sum_{ij} G_{ij}(\vec{x}-\vec{y}),
\end{equation}
where the two-vortex correlator $G_{ij}(\vec{x}-\vec{y})$ is
\begin{equation}
G_{ij}(\vec{x}-\vec{y})=\langle [ \delta(\vec{x}-\vec{x}^+_i) 
- \delta(\vec{x}-\vec{x}^-_i) ][
\delta(\vec{y}-\vec{x}^+_j) - \delta(\vec{y}-\vec{x}^-_j) ]\rangle,
\end{equation}
and the brackets $\langle\rangle$ indicate integration over the
positions $\vec{x}^\pm_i$.

As long as $i\ne j$, $G_{ij}(\vec{x}-\vec{y})=G_0(\vec{x}-\vec{y})$ is
independent of
$i$ and $j$. On the other hand, for $i=j$, we have
\begin{eqnarray}
G_{ii}(\vec{x}-\vec{y})
&=&\langle\delta(\vec{x}-\vec{x}^+_i)\delta(\vec{y}-\vec{x}^+_i)\rangle
\nonumber\\&&
+\langle\delta(\vec{x}-\vec{x}^-_i)\delta(\vec{y}-\vec{x}^-_i)\rangle
\nonumber\\&&
-\langle\delta(\vec{x}-\vec{x}^+_i)\delta(\vec{y}-\vec{x}^-_i)\rangle
\nonumber\\&&
-\langle\delta(\vec{x}-\vec{x}^-_i)\delta(\vec{y}-\vec{x}^+_i)\rangle.
\end{eqnarray}
The two last terms give contributions of order $N/A^2$, which we will ignore
but the two first give delta functions, so
\begin{equation}
G_{ii}(\vec{x}-\vec{y})=2\delta(\vec{x}-\vec{y})/A,
\end{equation}
where $A$ is the area of the film.

\begin{table}
\begin{tabular}{l|c|r}
\hline
electric charge & $e$ & $0.3$\\
coupling constants & $\alpha$ & $-0.25$\\
& $\beta$ & $0.18$\\
\hline
lattice size & $N^3$ & $512^3$\\
lattice spacing & $\delta x$ & $1.0$ \\
time step & $\delta t$ & $0.05$\\
\hline
initial temperature & $T_{\rm ini}$ & $10.0$\\
thermalization cycles & $N_{\rm th}$ & $16$\\
time evolved in each & $t_{\rm th}$ & $32.0$\\
\hline
\end{tabular}
\caption{
Parameter values
\labx{tab:params}
}
\end{table}

In total, we have
in Fourier space,
\begin{equation}
G(k)=\frac{4\pi^2}{e^2} \left[ (2N/A) + N(N-1) G_0(k) \right].
\end{equation}
The continuous-field limit is basically equivalent to taking
$N\rightarrow\infty$, and $2N/A=n$ is the number density of vortices,
so we find
\begin{equation} 
G(k)=\frac{4\pi^2n}{e^2} + G_{\rm cont}(k),
\end{equation}
where $G_{\rm cont}(k)$ is the two-point
function given by the continuous magnetic field.

Of course, the magnetic field is not completely localized, which means that
the delta functions spread over a finite distance. If we write the
Fourier transform of the magnetic field profile of a vortex as $G_1(k)$,
we have
\begin{equation} 
G(k)=\frac{4\pi^2n}{e^2}G_1(k) + G_{\rm cont}(k).
\end{equation}
We use the Gaussian ansatz in Eq.~(\ref{equ:fitGaussian})
for $G_{\rm cont}(k)$. 
In principle, $G_1(k)$ could be calculated by finding the static vortex
solution, but we simply parameterize it by an exponential, so that the
whole two-point function is
\begin{equation} 
G(k)=\frac{T_1k}{2}e^{-k/k_1}
+\frac{T_2k}{2}e^{-(k/k_2)^2}.
\labx{equ:fitfunction}
\end{equation}
We will use this form for fitting our numerical results.
The above discussion predicts that $k_1$ should not depend on 
the nature of the quench, because it is a property of the static
solution. The value of $T_1$ should be proportional to the number of vortices.

The linearized result in Eq.~(\ref{equ:exact2p}) 
is not of exactly the same form as Eq.~(\ref{equ:fitGaussian}), and therefore 
the parameters $T_2$ and $k_2$ will not be exactly the same as their
theoretical values in the fast-quench limit, but they should still be close
to them.

\section{Simulations}

\begin{figure}
\centerline{
\epsfig{file=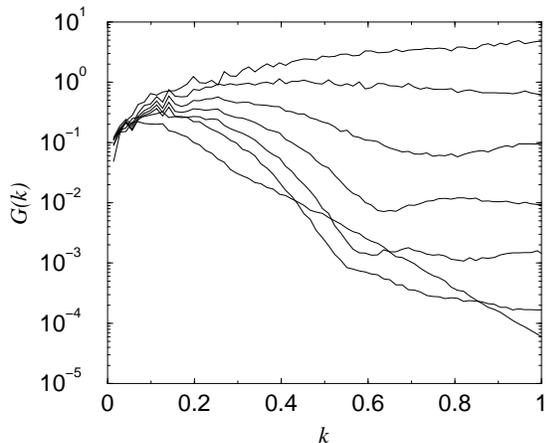,height=6cm}
}
\caption{
Time evolution of the two-point function at $\sigma=1.0$.
From top to bottom, the curves correspond to
$t=0$, $2$, $4$, $6$, $8$, $10$ and $20$.
\labx{fig:twopoint}
}
\end{figure}

\begin{figure}
\centerline{\epsfig{file=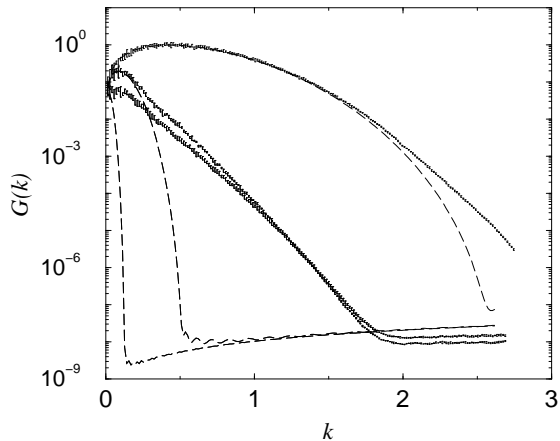,height=6cm}}
\caption{
The measured two point function $G(k)$, and the linearized prediction
calculated numerically from Eq.~(\ref{equ:exact2p}) [dashed lines]. 
From left to right, the curve pairs correspond to  $\sigma=0.25$, 
$\sigma=1.0$ and $\sigma=5.0$.
\labx{fig:fastquench}
}
\end{figure}

We tested our predictions by solving the
fully non-linear
equations of motion (\ref{equ:lateom}) numerically using the initial conditions
given by the partition function (\ref{equ:pf}). 
To do this, we defined the system
on a three-dimensional lattice, one slice of which corresponds to the film.
The details of the lattice implementation are presented in 
Appendix~\ref{app:lattice}.

In the preparation of the thermal initial conditions we employed the fact that
the energy functional $E_{\rm tot}$ in Eq.~(\ref{equ:latEtot}) is quadratic 
and diagonal in
${\mathbf E}$ and $\pi$. Therefore, their probability distribution is
Gaussian. The only complication is the Gauss law (\ref{equ:latGauss}), 
without which
the field values at different points would be uncorrelated.

To generate the initial conditions, we first drew random, uncorrelated values
for the component of $\pi$ that is parallel to $\psi$, i.e., 
$\pi_\psi=\psi ({\rm Re} \psi^*\pi)/(\psi^*\psi)$. 
This component does not appear in the
Gauss law (\ref{equ:latGauss}), 
and therefore it has the simple Gaussian probability
distribution
\begin{equation}
p\left[\pi_{\psi,(\vec{x})}\right]\propto \exp
\left[-\frac{\delta x^2}{T}\pi_{\psi,(\vec{x})}^2\right].
\end{equation}
Then, we went through all plaquettes and considered a change of the
electric field at all links around the plaquette by the same amount,
\begin{eqnarray}
{\mathbf E}_{i,(\mathbf x)}&\rightarrow&
{\mathbf E}_{i,(\mathbf x)}+\epsilon,\nonumber\\
{\mathbf E}_{j,(\mathbf x)+\hi}&\rightarrow&
{\mathbf E}_{j,(\mathbf x)+\hi}+\epsilon,\nonumber\\
{\mathbf E}_{i,(\mathbf x)+\hj}&\rightarrow&
{\mathbf E}_{i,(\mathbf x)+\hj}-\epsilon,\nonumber\\
{\mathbf E}_{j,(\mathbf x)}&\rightarrow&
{\mathbf E}_{j,(\mathbf x)}-\epsilon.
\end{eqnarray}
Such a change does not change the divergence of ${\mathbf E}$ and
therefore does not affect the Gauss law constraint. It would change the
energy by an amount that is at most quadratic in $\epsilon$,
\begin{equation}
\Delta E_{\rm tot}(\epsilon)=Q\epsilon^2+L\epsilon,
\end{equation}
where the constants $Q$ and $L$ depend on the field values at the neighbouring
links. Therefore the probability distribution for $\epsilon$ is
Gaussian,
\begin{equation}
p\left[\epsilon\right]\propto \exp
\left[-\frac{\Delta E_{\rm tot}(\epsilon)}{T}\right].
\end{equation}
It is straightforward to draw the value of $\epsilon$ from this
distribution.

Next we evolved the field configuration for time $t_{\rm th}$ using
the equations of motion (\ref{equ:lateom}), with zero conductivity. We 
repeated
this cycle of randomization and evolution steps $N_{\rm th}$ 
times, monitoring the 
evolution of the two-point function $G(k)$. When it reached the
equilibrium form given by the analytical calculation, we considered
the system to have thermalized.

We then solved the equations of motion (\ref{equ:lateom}) 
with the initial conditions
produced by the thermalization algorithm. We used a number of different
values for the conductivity $\sigma$ to test the dependence on the cooling
rate, and for each value we repeated the run several times using
different initial conditions from the same ensemble.
The values of the other parameters are shown in Table~\ref{tab:params}.

\section{Results}

\begin{figure}
\centerline{
\epsfig{file=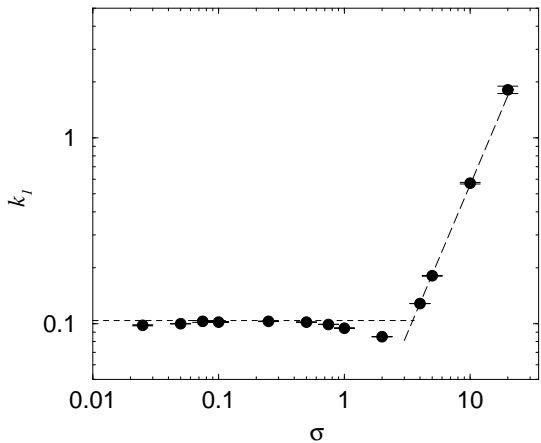,height=6cm}}
\caption{
The exponential cutoff scale $k_1$ as a function of $\sigma$. 
For $\sigma\rightarrow 0$ it approaches a constant value $k_1=0.104(1)$, which
characterizes the size of a static isolated vortex solution. 
At high $\sigma$, vortex
density is high and the system still away from equilibrium, and the
vortices are not well described by the static solution. In this regime,
we fit $k_1=0.0140(6)\sigma^{1.60(3)}$. In all other fits, we fixed
$k_1=0.1$.
\labx{fig:k1}
}
\end{figure}

\begin{figure}
\centerline{
\epsfig{file=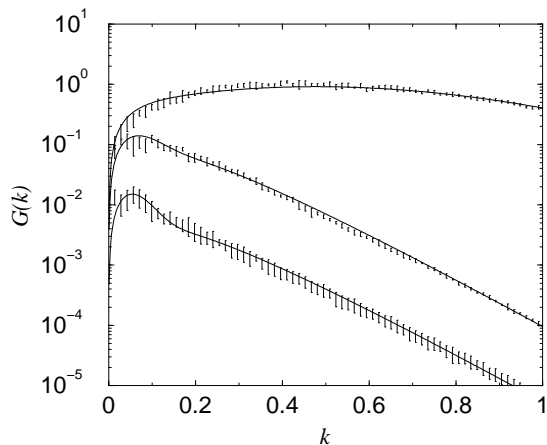,height=6cm}}
\caption{
Two point functions $G(k)$ together with fits of the form
(\ref{equ:fitfunction}). From top to bottom, the curves correspond to
$\sigma=5.0$, $0.5$ and $0.05$.
\labx{fig:fits}
}
\end{figure}

\begin{figure*}
\begin{tabular}{ll}
a)&b)\\
\epsfig{file=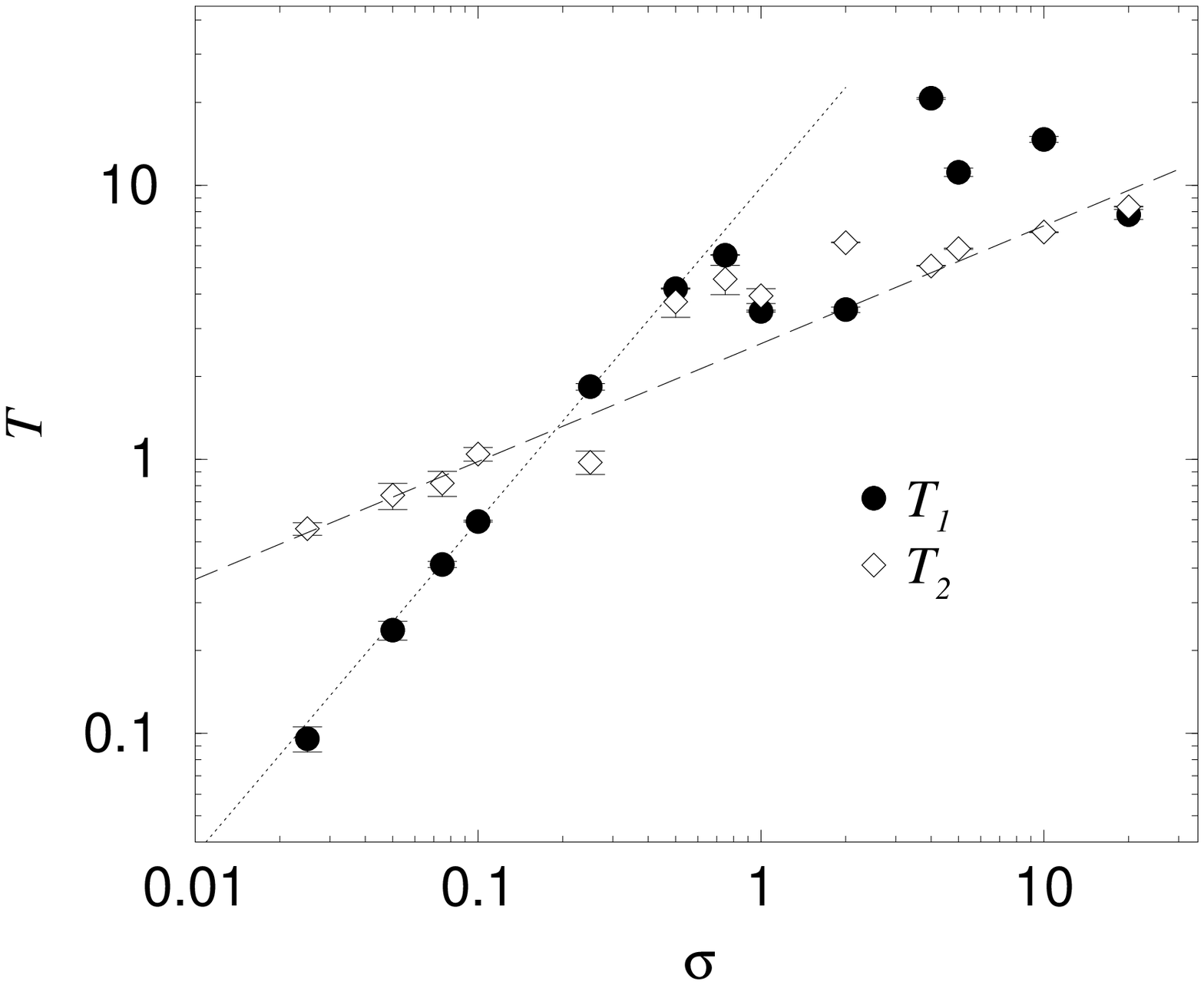,height=6cm}~~~~~~~~~~~~&
\epsfig{file=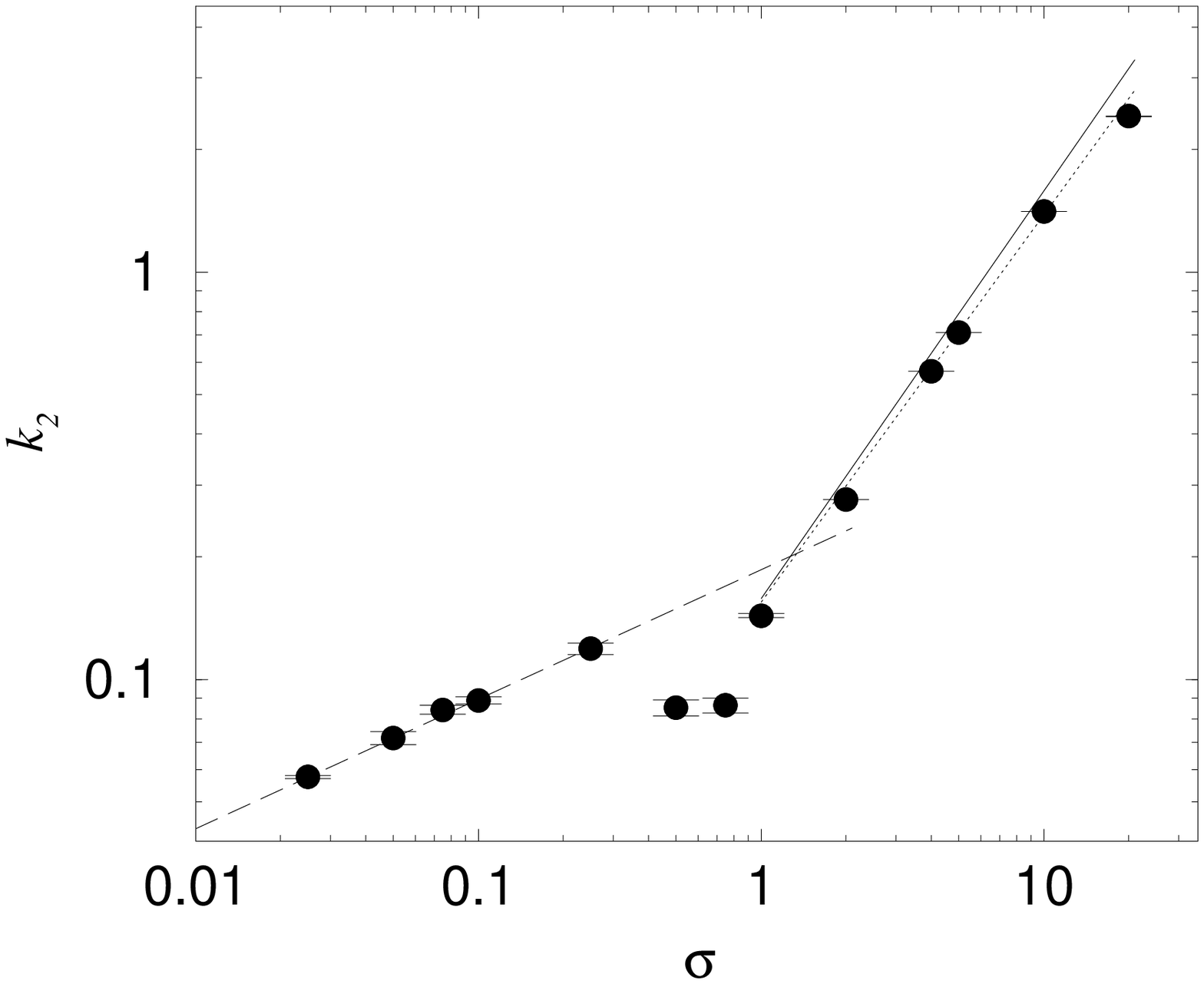,height=6cm}
\end{tabular}
\caption{
Fit parameters $T_1$, $T_2$ and $k_2$ as functions of $\sigma$.
(a) Coefficients $T_1$ (filled circles) and $T_2$ (empty diamonds)
of the exponential and Gaussian terms in Eq.~(\ref{equ:fitfunction})
respectively. The dotted line shows the power-law fit 
$T_1=9.76(9)\sigma^{1.218(6)}$ to the first six data points.
The dashed line shows a fit $T_2=2.6(5)\sigma^{0.43(6)}$ to the
first four data points.
(b) The critical wave-number $k_2$. 
The solid line shows the theoretical predition $k_2=\sqrt{\sigma/2t}$,
and the dashed and dotted lines are power-law fits 
$k_2=0.186(5)\sigma^{0.318(8)}$ and $k_2=0.115(7)\sigma^{0.95(3)}$
of the first five and last six data points, respectively.
\labx{fig:fitparams}
}
\end{figure*}

Our main aim was to test the flux trapping theory, and therefore
we measured the two-point function $G(k)$ of the magnetic field
fluctuations on the film. In Fig.~\ref{fig:twopoint}, we show the time 
evolution of $G(k)$ for $\sigma=1.0$. One can see that, as expected,
the long-wavelength modes freeze out to an high amplitude, whereas
short wavelengths are exponentially suppressed.
The plot also shows how the amplitude at around $0.3\lsim
k\lsim 0.8$
increases after $t\approx 10$, when the 
system enters the superconducting phase and 
the two-point function changes from
Eq.~(\ref{equ:fitGaussian}) to Eq.~(\ref{equ:fitfunction})
because of flux quantization.

To make it possible to compare the results for different values of $\sigma$,
we chose to carry out our measurements at time 
$t=20/\sigma$, so that the effective temperature is the same in each case.
Fig.~\ref{fig:fastquench} shows the measured two-points function at that time
for selected values of $\sigma$. They are compared with the results
for linearized friction-dominated freeze-out, calculated numerically
from the first line of Eq.~(\ref{equ:exact2p}).
The plot shows that the linearized result works very well in fast
quenches. For $\sigma\lsim 1$, one can see a clear discrepancy, which
is a sign that the non-linear effects have become important.
The difference is also
partly due to the contribution $G_1(k)$ from flux quantization.

We fitted 
the two-point functions with the ansatz in Eq.~(\ref{equ:fitfunction}).
In Fig.~\ref{fig:k1}, we show $k_1$ as a function of $\sigma$.
As expected, its value is independent of $\sigma$ in slow quenches.
At higher $\sigma$, the vortex density becomes higher and non-equilibrium
effects more important, and therefore it is not surprising that
$k_1$ starts to grow.
This can be interpreted as the size of the vortex getting smaller
as more of them are packed in the same area.

To improve the accuracy of the fit parameters in subsequent fits,
we fixed $k_1$ to 0.1.
Some examples of these fits are shown in Fig.~\ref{fig:fits}.
The plateau at very short wavelengths, which corresponds to modes still
in thermal equilibrium was excluded from the fits. 

In Fig.~\ref{fig:fitparams} we show the fit parameters $T_1$, $T_2$ and $k_2$
as functions of $\sigma$. The errors were estimated using
the bootstrap method, and contain only the statistical error.
We did not attempt to estimate the systematic error in any measurement due to
the choice of the fitting function.

In the high-$\sigma$
limit, the two-point function should be given by Eq.~(\ref{equ:exact2p}),
and to the extent that it can be approximated by a Gaussian, we expect
$T_2\approx T=10$ and $k_2\approx \sqrt{\sigma/2t}$.
This is confirmed by the measurements.
In this limit the values of $T_1$ show significant scattering, but this
understandable, because $G(k)$ is dominated by the Gaussian term and we
had also fixed $k_1$ to a very different value from the best fit.

In the opposite limit of low cooling rates,
the parameters seem to be well described
by power laws,
\begin{eqnarray}
T_1&\sim& 9.76(9)\sigma^{1.218(6)},
\nonumber\\
T_2&\sim& 2.6(5)\sigma^{0.43(6)},
\nonumber\\
k_2&\sim& 
0.186(5)\sigma^{0.318(8)}.
\label{equ:asymp}
\end{eqnarray}
The deviation from the linear prediction $k_2\approx\sqrt{\sigma/2t}$
is a sign of non-linear dynamics. It is interesting to note,
but possibly a coincidence, that
the behaviour of $T_2$ appears to agree with the same power law even at
$\sigma\approx 10$.

\begin{figure}
\centerline{
\epsfig{file=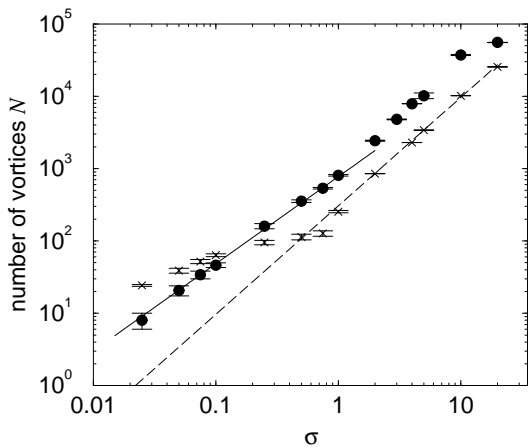,height=6cm}}
\caption{
Vortex number measured at $\sigma t=20$ as a function of $\sigma$.
The dashed line shows the theoretical prediction (\ref{equ:npredfast})
for fast quenches, and the solid line a 
a power-law fit
$N=772(18)\sigma^{1.20(2)}$ of data at $\sigma\le 0.5$.
The crosses show the prediction (\ref{equ:npredslow}) for slow quenches
based on the fit parameters $T_2$ and $k_2$.
\labx{fig:vortexnumber}
}
\end{figure}

Fig.~\ref{fig:vortexnumber} shows the number of vortices plus antivortices
measured at time $\sigma t=20$. The agreement with the analytical 
prediction Eq.~(\ref{equ:npredfast}) shown by the dashed line is
good for fast quenches $\sigma\gsim 1$, apart from a constant factor of
about 2.
This indicates that vortices are predominantly formed by
flux trapping.

For $\sigma\lsim 1$, we can fit the data very well with a power
law $N=772(18)\sigma^{1.20(2)}$.
This is compatible with the expectation that $T_1\propto N$.
We have also plotted the
theoretical prediction in 
Eq.~(\ref{equ:npredslow}) calculated using the fitted parameter values. 
As the plot shows, it gives the correct order of magnitude, although it
does not agree perfectly for slow quenches.
The prediction seems to suggest a different power-law behaviour
from the observed one.

\begin{figure}
\centerline{
\epsfig{file=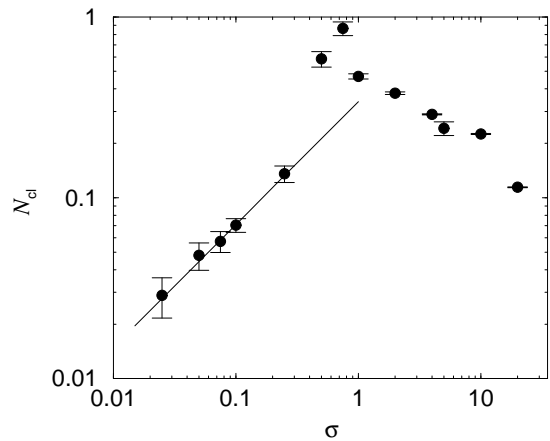,height=6cm}
}
\caption{
The number of vortices in each cluster calculated from the measured vortex
density and the fit parameter $k_2$.
\labx{fig:Ncl}
}
\end{figure}

As Fig.~\ref{fig:twopoint} shows, the fluctuations
have frozen out, but it is possible that their amplitude is not
high enough to actually dominate the process of vortex formation.
Indeed, if one estimates the typical number of vortices in a cluster
using Eq.~(\ref{equ:Ncl})
one finds it decreases rapidly at low $\sigma$ (see Fig.~\ref{fig:Ncl}).
In principle, the flux trapping mechanism requires this number to
be greater than one.
This suggest that the slow quenches could possibly be 
better described by the Kibble-Zurek 
mechanism.

\begin{figure}
\centerline{
\epsfig{file=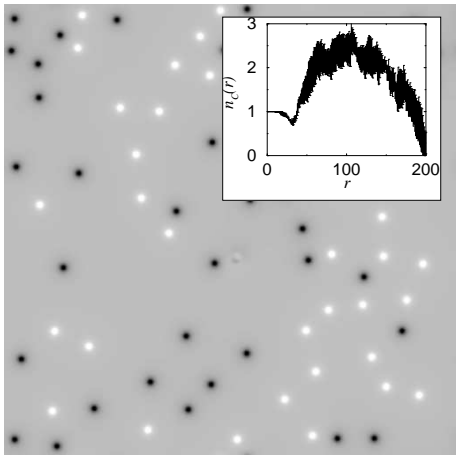,height=6cm}
}
\caption{
The magnetic field configuration on the film at time $t=400$ in a run with
$T=1000$ and $\sigma=0.25$.
The curve in the inset shows the quantity $n_C(r)$,
which measures clustering.
\labx{fig:snapshots}}
\end{figure}

It is possible to increase $N_{\rm cl}$ by using different parameter values,
and we carried out one set of runs with $T=1000$.
The clustering is indeed evident in the 
snapshot of the magnetic field at time $t=400$ in a run with 
$\sigma=0.25$ shown in 
Fig.~\ref{fig:snapshots}. In the same figure, we have also plotted the 
quantity,
$n_C(r)$ which measures the average winding number around a circle of radius
$r$ centered at a vortex.\cite{Hindmarsh:2000kd}
Clustering is indicated by values greater than
one, and the maximum value gives a measure of the typical number of
vortices in a cluster.

\section{Conclusions}

In fast quenches $\sigma\gg 1$, 
the two-point function $G(k)$ of the magnetic field 
fluctuations is consistent with the predictions of the linearized theory.
This allows us to calculate the number of vortices produced by flux
trapping, and this prediction agrees very well with the measured values.
This provides strong support for the flux trapping scenario of
vortex formation. Furthermore, the spatial distribution of vortices 
shows the tell-tale signal of vortex clusters.

It should be kept in mind that in fast quenches, the relevant 
degrees of freedom are overdamped. This is not necessarily a problem,
because it can be given a physical interpretation as a relatively high
conductivity. Furthermore, overdamped dynamics is commonly used
in studies of superconductors and vortex formation.\cite{Stephens:2001fv}
Nevertheless, for many applications underdamped dynamics would probably
be better. Ideally one should carry out the simulation without introducing 
damping at all,\cite{Hindmarsh:2000kd,Hindmarsh:2001vp} but
even in our case, the relevant degrees of freedom are underdamped in
slow quenches.

The underdamped nature of dynamics makes it much more difficult to
describe slow quenches theoretically,
and therefore the predictions are not as robust. In spite of this, our 
results show a reasonably good agreement between theory and simulation,
although the scaling of the vortex number with the cooling rate $\sigma$
does not seem to agree. This may be an indication that the amplitude
of trapped fluctuations is so low that vortex formation can be better 
described by the Kibble-Zurek mechanism.

Simulations with a higher initial temperature support this picture, 
and show clear vortex clustering even in slow quenches. This
is an unmistakable sign of flux trapping 

To understand better how vortices form when the amplitude is not
high enough for flux trapping,
it would be important to derive theoretical predictions of the 
Kibble-Zurek scenario and compare them with our results.
In particular, one must understand how a weak background magnetic field
biases the Kibble-Zurek mechanism.

Our results show that under the right conditions, the electromagnetic
field plays an important role in defect formation. It remains to be
seen if these conditions are satisfied in actual superconductor
experiments. Likewise, the cosmological consequences of this remain
largely unexplored.

\begin{acknowledgments}
This work was supported by Churchill College, Cambridge (A.R.), 
Cambridge European Trust and EPSRC (M.D.).
We also acknowledge support from the European Science Foundation,
through the COSLAB (Cosmology in the Laboratory) programme.
This research was conducted in cooperation with SGI/Intel          
utilising the Altix 3700 supercomputer.
\end{acknowledgments}

\appendix
\section{Lattice discretization\labx{app:lattice}}

To write down the discretized equation of motion, we define the 
forward and backward derivatives
\begin{equation}
\Delta^\pm_i f_{({\mathbf x})}=
\pm\delta x^{-1}\left(f_{({\mathbf x}\pm\hi)}-f_{({\mathbf x})}\right),
\end{equation}
where $\delta x$ is the lattice spacing and $\hi$ is a unit vector in
the $i$ direction.
Similarly, we define the time derivative
\begin{equation}
\Delta_t f_{(t)}
=\delta t^{-1}\left(f_{(t)}-f_{(t-\delta t)}\right),
\end{equation}
where $\delta t$ is the time step.
Using the
link variable
\begin{equation}
\vec{U}_{i}=\exp\left(ie\delta x \vec{A}_{i}\right),
\end{equation}
we also define the corresponding covariant derivatives on the film
\begin{eqnarray}
\vec{D}^+_i\psi_{(\vec{x})}&=&\delta x^{-1}\left(
\vec{U}_{i,(\vec{x})}\psi_{(\vec{x}+\hi)}
-\psi_{(\vec{x})}\right),\nonumber\\
\vec{D}^-_i\psi_{(\vec{x})}&=&\delta x^{-1}\left(
\psi_{(\vec{x})}-
\vec{U}^*_{i,(\vec{x}-\hi)}\psi_{(\vec{x}-\hi)}\right)
.
\end{eqnarray}

The discretized equations of motion are
\begin{eqnarray}
\Delta_t {\mathbf A}_{i,(t,{\mathbf x})}&=&-
{\mathbf E}_{i,(t-\delta t,{\mathbf x})},
\nonumber\\
\Delta_t\psi_{(t,\vec{x})}&=&\pi_{(t-\delta t,\vec{x})},
\nonumber\\
\Delta_t {\mathbf E}_{i,(t,{\mathbf x})}&=&
\sum_{jklm}
\epsilon_{ijk}\epsilon_{klm}
\Delta^-_j\Delta^+_l{\mathbf A}_{m,(t,{\mathbf x})}
-\sigma {\mathbf E}_{i,(t,{\mathbf x})}\nonumber\\&&
-\frac{2 e}{\delta x} 
\delta_z {\rm Im}\psi^*_{(t,\vec{x})}\vec{D}^+_i\psi_{(t,\vec{x})},
\nonumber\\
\Delta_t\pi_{(t,\vec{x})}&=&
\sum_i
\vec{D}_i^-\vec{D}_i^+\psi_{(t,\vec{x})}
-\sigma\pi_{(t,\vec{x})}
\nonumber\\&&
-\alpha\psi_{(t,\vec{x})}
-\beta|\psi_{(t,\vec{x})}|^2\psi_{(t,\vec{x})},
\labx{equ:lateom}
\end{eqnarray}
where the summation over directions
goes from $1$ to either $2$ or $3$ depending
on the context.

The lattice version of the Gauss law is
\begin{equation}
\sum_i\Delta^-_i{\mathbf E}_{i,(t,{\mathbf x})}=
\frac{2e}{\delta x}\delta_z{\rm Im}\psi^*_{(t,\vec{x})}
\pi_{(t,\vec{x})}.
\labx{equ:latGauss}
\end{equation}
This equality is exactly conserved by the lattice equations of motion
and is a constraint the initial field configuration must
satisfy.

The lattice version of the energy functional (\ref{equ:totalenergy}) is
\begin{eqnarray}
E_{\rm tot}&=&\frac{1}{2}\sum_{{\mathbf x},i}\delta x^3\left[
{\mathbf E}_{i}^2
+\sum_{jk}\left(\epsilon_{ijk}\Delta_j^+{\mathbf A}_k\right)^2
\right]\nonumber\\
&&+\sum_{\vec x}\delta x^2\Bigl[
\pi^*\pi
-\frac{2}{\delta x^2}\sum_i{\rm Re}\psi_{(\vec x)}^*
\vec{U}_{i,(\vec x)}\psi_{(\vec x+\hi)}\nonumber\\
&&~~~~~~~~~~~
+\left(\alpha+\frac{4}{\delta x^2}\right)|\psi|^2+\frac{\beta}{2}|\psi|^4
\Bigr].
\labx{equ:latEtot}
\end{eqnarray}

It should be noted that both the equations of motion (\ref{equ:lateom}) 
and the energy functional (\ref{equ:latEtot}) 
are gauge invariant. In this non-compact
lattice formulation, the gauge group is, in fact, $\mathbb R$ rather than
U(1), which has the advantage that vortices are topologically stable even
on lattice.

In the actual simulations, we used a finite lattice with periodic
boundary conditions in all three directions. This has the consequence
that the net magnetic flux through the film vanishes.

\bibliography{filmsimu}

\begin{thebibliography}{29}
\expandafter\ifx\csname natexlab\endcsname\relax\def\natexlab#1{#1}\fi
\expandafter\ifx\csname bibnamefont\endcsname\relax
  \def\bibnamefont#1{#1}\fi
\expandafter\ifx\csname bibfnamefont\endcsname\relax
  \def\bibfnamefont#1{#1}\fi
\expandafter\ifx\csname citenamefont\endcsname\relax
  \def\citenamefont#1{#1}\fi
\expandafter\ifx\csname url\endcsname\relax
  \def\url#1{\texttt{#1}}\fi
\expandafter\ifx\csname urlprefix\endcsname\relax\def\urlprefix{URL }\fi
\providecommand{\bibinfo}[2]{#2}
\providecommand{\eprint}[2][]{\url{#2}}

\bibitem[{\citenamefont{Carmi and Polturak}(1999)}]{carmi:7595}
\bibinfo{author}{\bibfnamefont{R.}~\bibnamefont{Carmi}} \bibnamefont{and}
  \bibinfo{author}{\bibfnamefont{E.}~\bibnamefont{Polturak}},
  \bibinfo{journal}{Phys. Rev. B} \textbf{\bibinfo{volume}{60}},
  \bibinfo{pages}{7595} (\bibinfo{year}{1999}).

\bibitem[{\citenamefont{Carmi et~al.}(2000)\citenamefont{Carmi, Polturak, and
  Koren}}]{carmi:4966}
\bibinfo{author}{\bibfnamefont{R.}~\bibnamefont{Carmi}},
  \bibinfo{author}{\bibfnamefont{E.}~\bibnamefont{Polturak}}, \bibnamefont{and}
  \bibinfo{author}{\bibfnamefont{G.}~\bibnamefont{Koren}},
  \bibinfo{journal}{Phys. Rev. Lett.} \textbf{\bibinfo{volume}{84}},
  \bibinfo{pages}{4966} (\bibinfo{year}{2000}).

\bibitem[{\citenamefont{Kavoussanaki et~al.}(2000)\citenamefont{Kavoussanaki,
  Monaco, and Rivers}}]{Kavoussanaki:2000tj}
\bibinfo{author}{\bibfnamefont{E.}~\bibnamefont{Kavoussanaki}},
  \bibinfo{author}{\bibfnamefont{R.}~\bibnamefont{Monaco}}, \bibnamefont{and}
  \bibinfo{author}{\bibfnamefont{R.~J.} \bibnamefont{Rivers}},
  \bibinfo{journal}{Phys. Rev. Lett.} \textbf{\bibinfo{volume}{85}},
  \bibinfo{pages}{3452} (\bibinfo{year}{2000}), \eprint{cond-mat/0005145}.

\bibitem[{\citenamefont{Monaco et~al.}(2002)\citenamefont{Monaco, Mygind, and
  Rivers}}]{monaco:080603}
\bibinfo{author}{\bibfnamefont{R.}~\bibnamefont{Monaco}},
  \bibinfo{author}{\bibfnamefont{J.}~\bibnamefont{Mygind}}, \bibnamefont{and}
  \bibinfo{author}{\bibfnamefont{R.~J.} \bibnamefont{Rivers}},
  \bibinfo{journal}{Phys. Rev. Lett.} \textbf{\bibinfo{volume}{89}},
  \bibinfo{eid}{080603} (\bibinfo{year}{2002}).

\bibitem[{\citenamefont{Monaco et~al.}(2003)\citenamefont{Monaco, Mygind, and
  Rivers}}]{monaco:104506}
\bibinfo{author}{\bibfnamefont{R.}~\bibnamefont{Monaco}},
  \bibinfo{author}{\bibfnamefont{J.}~\bibnamefont{Mygind}}, \bibnamefont{and}
  \bibinfo{author}{\bibfnamefont{R.~J.} \bibnamefont{Rivers}},
  \bibinfo{journal}{Phys. Rev. B} \textbf{\bibinfo{volume}{67}},
  \bibinfo{eid}{104506} (\bibinfo{year}{2003}).

\bibitem[{\citenamefont{Kirtley et~al.}(2003)\citenamefont{Kirtley, Tsuei, and
  Tafuri}}]{kirtley:257001}
\bibinfo{author}{\bibfnamefont{J.~R.} \bibnamefont{Kirtley}},
  \bibinfo{author}{\bibfnamefont{C.~C.} \bibnamefont{Tsuei}}, \bibnamefont{and}
  \bibinfo{author}{\bibfnamefont{F.}~\bibnamefont{Tafuri}},
  \bibinfo{journal}{Phys. Rev. Lett.} \textbf{\bibinfo{volume}{90}},
  \bibinfo{eid}{257001} (\bibinfo{year}{2003}).

\bibitem[{\citenamefont{Maniv et~al.}(2003)\citenamefont{Maniv, Polturak, and
  Koren}}]{maniv:197001}
\bibinfo{author}{\bibfnamefont{A.}~\bibnamefont{Maniv}},
  \bibinfo{author}{\bibfnamefont{E.}~\bibnamefont{Polturak}}, \bibnamefont{and}
  \bibinfo{author}{\bibfnamefont{G.}~\bibnamefont{Koren}},
  \bibinfo{journal}{Phys. Rev. Lett.} \textbf{\bibinfo{volume}{91}},
  \bibinfo{eid}{197001} (\bibinfo{year}{2003}).

\bibitem[{\citenamefont{Kirtley et~al.}(2004)\citenamefont{Kirtley, Tsuei,
  Tafuri, Medaglia, Orgiani, and Balestrino}}]{Kir04}
\bibinfo{author}{\bibfnamefont{J.~R.} \bibnamefont{Kirtley}},
  \bibinfo{author}{\bibfnamefont{C.~C.} \bibnamefont{Tsuei}},
  \bibinfo{author}{\bibfnamefont{F.}~\bibnamefont{Tafuri}},
  \bibinfo{author}{\bibfnamefont{P.~G.} \bibnamefont{Medaglia}},
  \bibinfo{author}{\bibfnamefont{P.}~\bibnamefont{Orgiani}}, \bibnamefont{and}
  \bibinfo{author}{\bibfnamefont{G.}~\bibnamefont{Balestrino}},
  \bibinfo{journal}{Supercond. Sci. Technol.} \textbf{\bibinfo{volume}{17}},
  \bibinfo{pages}{S217} (\bibinfo{year}{2004}).

\bibitem[{\citenamefont{Bauerle et~al.}(1996)}]{Bau+96}
\bibinfo{author}{\bibfnamefont{C.}~\bibnamefont{Bauerle}} \bibnamefont{et~al.},
  \bibinfo{journal}{Nature} \textbf{\bibinfo{volume}{382}},
  \bibinfo{pages}{332} (\bibinfo{year}{1996}).

\bibitem[{\citenamefont{Ruutu et~al.}(1996)}]{Ruutu:1996qz}
\bibinfo{author}{\bibfnamefont{V.~M.~H.} \bibnamefont{Ruutu}}
  \bibnamefont{et~al.}, \bibinfo{journal}{Nature}
  \textbf{\bibinfo{volume}{382}}, \bibinfo{pages}{334} (\bibinfo{year}{1996}),
  \eprint{cond-mat/9512117}.

\bibitem[{\citenamefont{Dodd et~al.}(1998)\citenamefont{Dodd, Hendry, Lawson,
  McClintock, and Williams}}]{Dod+98}
\bibinfo{author}{\bibfnamefont{M.~E.} \bibnamefont{Dodd}},
  \bibinfo{author}{\bibfnamefont{P.~C.} \bibnamefont{Hendry}},
  \bibinfo{author}{\bibfnamefont{N.~S.} \bibnamefont{Lawson}},
  \bibinfo{author}{\bibfnamefont{P.~V.~E.} \bibnamefont{McClintock}},
  \bibnamefont{and} \bibinfo{author}{\bibfnamefont{C.~D.~H.}
  \bibnamefont{Williams}}, \bibinfo{journal}{Phys. Rev. Lett.}
  \textbf{\bibinfo{volume}{81}}, \bibinfo{pages}{3703} (\bibinfo{year}{1998}).

\bibitem[{\citenamefont{Chuang et~al.}(1991)\citenamefont{Chuang, Durrer,
  Turok, and Yurke}}]{Chu+91}
\bibinfo{author}{\bibfnamefont{I.}~\bibnamefont{Chuang}},
  \bibinfo{author}{\bibfnamefont{R.}~\bibnamefont{Durrer}},
  \bibinfo{author}{\bibfnamefont{N.}~\bibnamefont{Turok}}, \bibnamefont{and}
  \bibinfo{author}{\bibfnamefont{B.}~\bibnamefont{Yurke}},
  \bibinfo{journal}{Science} \textbf{\bibinfo{volume}{251}},
  \bibinfo{pages}{1336} (\bibinfo{year}{1991}).

\bibitem[{\citenamefont{Bowick et~al.}(1994)\citenamefont{Bowick, Chandar,
  Schiff, and Srivastava}}]{Bow+94}
\bibinfo{author}{\bibfnamefont{M.~J.} \bibnamefont{Bowick}},
  \bibinfo{author}{\bibfnamefont{L.}~\bibnamefont{Chandar}},
  \bibinfo{author}{\bibfnamefont{E.~A.} \bibnamefont{Schiff}},
  \bibnamefont{and} \bibinfo{author}{\bibfnamefont{A.~M.}
  \bibnamefont{Srivastava}}, \bibinfo{journal}{Science}
  \textbf{\bibinfo{volume}{263}}, \bibinfo{pages}{943} (\bibinfo{year}{1994}),
  \eprint{hep-ph/9208233}.

\bibitem[{\citenamefont{Digal et~al.}(1999)\citenamefont{Digal, Ray, and
  Srivastava}}]{Dig98}
\bibinfo{author}{\bibfnamefont{S.}~\bibnamefont{Digal}},
  \bibinfo{author}{\bibfnamefont{R.}~\bibnamefont{Ray}}, \bibnamefont{and}
  \bibinfo{author}{\bibfnamefont{A.~M.} \bibnamefont{Srivastava}},
  \bibinfo{journal}{Phys. Rev. Lett.} \textbf{\bibinfo{volume}{83}},
  \bibinfo{pages}{5030} (\bibinfo{year}{1999}), \eprint{hep-ph/9805502}.

\bibitem[{\citenamefont{Ray and Srivastava}(2004)}]{Ray04}
\bibinfo{author}{\bibfnamefont{R.}~\bibnamefont{Ray}} \bibnamefont{and}
  \bibinfo{author}{\bibfnamefont{A.~M.} \bibnamefont{Srivastava}},
  \bibinfo{journal}{Phys. Rev.} \textbf{\bibinfo{volume}{D69}},
  \bibinfo{pages}{103525} (\bibinfo{year}{2004}), \eprint{hep-ph/0110165}.

\bibitem[{\citenamefont{Kibble}(1976)}]{Kibble:1976sj}
\bibinfo{author}{\bibfnamefont{T.~W.~B.} \bibnamefont{Kibble}},
  \bibinfo{journal}{J. Phys.} \textbf{\bibinfo{volume}{A9}},
  \bibinfo{pages}{1387} (\bibinfo{year}{1976}).

\bibitem[{\citenamefont{Kibble}(1980)}]{Kibble:1980mv}
\bibinfo{author}{\bibfnamefont{T.~W.~B.} \bibnamefont{Kibble}},
  \bibinfo{journal}{Phys. Rept.} \textbf{\bibinfo{volume}{67}},
  \bibinfo{pages}{183} (\bibinfo{year}{1980}).

\bibitem[{\citenamefont{Zurek}(1996)}]{Zurek:1996sj}
\bibinfo{author}{\bibfnamefont{W.~H.} \bibnamefont{Zurek}},
  \bibinfo{journal}{Phys. Rept.} \textbf{\bibinfo{volume}{276}},
  \bibinfo{pages}{177} (\bibinfo{year}{1996}), \eprint{cond-mat/9607135}.

\bibitem[{\citenamefont{Hindmarsh and Rajantie}(2000)}]{Hindmarsh:2000kd}
\bibinfo{author}{\bibfnamefont{M.}~\bibnamefont{Hindmarsh}} \bibnamefont{and}
  \bibinfo{author}{\bibfnamefont{A.}~\bibnamefont{Rajantie}},
  \bibinfo{journal}{Phys. Rev. Lett.} \textbf{\bibinfo{volume}{85}},
  \bibinfo{pages}{4660} (\bibinfo{year}{2000}), \eprint{cond-mat/0007361}.

\bibitem[{\citenamefont{Rajantie}(2002)}]{Rajantie:2001ps}
\bibinfo{author}{\bibfnamefont{A.}~\bibnamefont{Rajantie}},
  \bibinfo{journal}{Int. J. Mod. Phys.} \textbf{\bibinfo{volume}{A17}},
  \bibinfo{pages}{1} (\bibinfo{year}{2002}), \eprint{hep-ph/0108159}.

\bibitem[{\citenamefont{Vilenkin and Shellard}(1994)}]{VilShel}
\bibinfo{author}{\bibfnamefont{A.}~\bibnamefont{Vilenkin}} \bibnamefont{and}
  \bibinfo{author}{\bibfnamefont{E.}~\bibnamefont{Shellard}},
  \emph{\bibinfo{title}{Cosmic strings and other topological defects}}
  (\bibinfo{publisher}{Cambridge University Press}, \bibinfo{year}{1994}).

\bibitem[{\citenamefont{Dvali and Vilenkin}(2004)}]{Dvali:2003zj}
\bibinfo{author}{\bibfnamefont{G.}~\bibnamefont{Dvali}} \bibnamefont{and}
  \bibinfo{author}{\bibfnamefont{A.}~\bibnamefont{Vilenkin}},
  \bibinfo{journal}{JCAP} \textbf{\bibinfo{volume}{0403}}, \bibinfo{pages}{010}
  (\bibinfo{year}{2004}), \eprint{hep-th/0312007}.

\bibitem[{\citenamefont{Sazhin et~al.}(2003)}]{Sazhin:2003cp}
\bibinfo{author}{\bibfnamefont{M.}~\bibnamefont{Sazhin}} \bibnamefont{et~al.},
  \bibinfo{journal}{Mon. Not. Roy. Astron. Soc.}
  \textbf{\bibinfo{volume}{343}}, \bibinfo{pages}{353} (\bibinfo{year}{2003}),
  \eprint{astro-ph/0302547}.

\bibitem[{\citenamefont{Schild et~al.}(2004)\citenamefont{Schild, Masnyak,
  Hnatyk, and Zhdanov}}]{Schild:2004uv}
\bibinfo{author}{\bibfnamefont{R.~E.} \bibnamefont{Schild}},
  \bibinfo{author}{\bibfnamefont{I.~S.} \bibnamefont{Masnyak}},
  \bibinfo{author}{\bibfnamefont{B.~I.} \bibnamefont{Hnatyk}},
  \bibnamefont{and} \bibinfo{author}{\bibfnamefont{V.~I.}
  \bibnamefont{Zhdanov}} (\bibinfo{year}{2004}), \eprint{astro-ph/0406434}.

\bibitem[{\citenamefont{Sazhin et~al.}(2004)}]{Sazhin:2004fv}
\bibinfo{author}{\bibfnamefont{M.~V.} \bibnamefont{Sazhin}}
  \bibnamefont{et~al.} (\bibinfo{year}{2004}), \eprint{astro-ph/0406516}.

\bibitem[{\citenamefont{Nielsen and Olesen}(1973)}]{Nielsen:1973cs}
\bibinfo{author}{\bibfnamefont{H.~B.} \bibnamefont{Nielsen}} \bibnamefont{and}
  \bibinfo{author}{\bibfnamefont{P.}~\bibnamefont{Olesen}},
  \bibinfo{journal}{Nucl. Phys.} \textbf{\bibinfo{volume}{B61}},
  \bibinfo{pages}{45} (\bibinfo{year}{1973}).

\bibitem[{\citenamefont{Kibble and Rajantie}(2003)}]{Kib03}
\bibinfo{author}{\bibfnamefont{T.~W.~B.} \bibnamefont{Kibble}}
  \bibnamefont{and} \bibinfo{author}{\bibfnamefont{A.}~\bibnamefont{Rajantie}},
  \bibinfo{journal}{Phys. Rev.} \textbf{\bibinfo{volume}{B68}},
  \bibinfo{pages}{174512} (\bibinfo{year}{2003}), \eprint{cond-mat/0306633}.

\bibitem[{\citenamefont{Stephens et~al.}(2002)\citenamefont{Stephens,
  Bettencourt, and Zurek}}]{Stephens:2001fv}
\bibinfo{author}{\bibfnamefont{G.~J.} \bibnamefont{Stephens}},
  \bibinfo{author}{\bibfnamefont{L.~M.~A.} \bibnamefont{Bettencourt}},
  \bibnamefont{and} \bibinfo{author}{\bibfnamefont{W.~H.} \bibnamefont{Zurek}},
  \bibinfo{journal}{Phys. Rev. Lett.} \textbf{\bibinfo{volume}{88}},
  \bibinfo{pages}{137004} (\bibinfo{year}{2002}), \eprint{cond-mat/0108127}.

\bibitem[{\citenamefont{Hindmarsh and Rajantie}(2001)}]{Hindmarsh:2001vp}
\bibinfo{author}{\bibfnamefont{M.}~\bibnamefont{Hindmarsh}} \bibnamefont{and}
  \bibinfo{author}{\bibfnamefont{A.}~\bibnamefont{Rajantie}},
  \bibinfo{journal}{Phys. Rev.} \textbf{\bibinfo{volume}{D64}},
  \bibinfo{pages}{065016} (\bibinfo{year}{2001}), \eprint{hep-ph/0103311}.

\end{thebibliography}
\end{document}